# Airborne liquid marble: Evaporation dynamics of liquid marble in acoustic levitation

Keigo Noro[a], Xiao Ma[b] and Koji Hasegawa*[c]

Liquid marbles (LMs)—droplets encapsulated by hydrophobic particles—allow for non-wetting manipulation and containerless handling, making them well-suited for applications such as microreactors. When integrated with acoustic levitation, the LMs serve as contact-free reaction platforms. However, the complex behaviours associated with acoustic fields, including deformation, internal flow, and evaporation, remain insufficiently understood. This study investigates the evaporation dynamics of acoustically levitated LMs. Simultaneous visualisation of the droplet morphology and surrounding vapour concentration fields was achieved using backlighting and the Background-oriented Schlieren (BOS) method. At intermediate relative humidity (RH ≈ 40%), the evaporation rates and vapour distributions of LMs closely resembled those of pure water droplets, indicating that the particle shell exerts minimal influence under these conditions. Conversely, significant differences were observed at high- and low-humidity, attributable to the evaporation resistance and interfacial properties introduced by the hydrophobic particle layers. Furthermore, although pure water droplets maintained a quasi-spherical shape, LMs demonstrated progressive flattening over time, suggesting mechanical constraints imposed by the particle shell. These findings provide new insights into the coupled evaporation and deformation behaviours of LMs in acoustic fields, informing the design of contamination-free microfluidic and microreactor systems.

## 1. Introduction

Recently, liquid marbles (LMs), which are non-adhesive droplets encapsulated by a shell of hydrophobic particles, have garnered significant attention across a broad spectrum of scientific disciplines [1-5]. LMs were first systematically described by Aussillous and Quéré in 2001, who demonstrated that rolling a water droplet on a hydrophobic powder bed produces a stable, non-wetting structure in which the liquid core is completely isolated from the environment by a porous particle shell [1]. This distinctive architecture, characterised by low surface friction, high mobility, and gas permeability through the shell, underpins the versatility across diverse applications [2-4].

In the pharmaceutical sector, LMs have been investigated as precursors for hollow granules and microcapsules [6], as well as components of Pickering emulsions [7], providing a surfactant-free platform for encapsulating hydrophobic active ingredients and enabling controlled or targeted drug release [8, 9]. In the medical field, LMs have been utilised as miniature bioreactors for biological assays, such as rapid blood typing [10, 11], and as three-dimensional cell culture platforms. Notably, cancer cells cultured within LMs spontaneously aggregate into spheroids that more closely replicate the architecture of solid tumours compared with conventional two-dimensional cultures [12, 13]. This capability has been further extended to the formation of embryoid bodies from embryonic stem cells and their subsequent cardiogenesis, as well as to the cryopreservation of mammalian cells [14, 15], highlighting the potential of LMs in tissue engineering and regenerative medicine. In chemical and analytical applications, LMs serve as microscale reactors: their enclosed volume supports a variety of chemical transformations [16], whereas the gas-permeable shell allows indicator solutions within the marble to detect target gases, such as ammonia, through visible colour changes [16, 17].

The inherent resistance of LMs to wetting and adhesion provides a significant advantage, allowing for the manipulation of microdroplets in a containerless environment [2, 5]. Therefore, the integration of LMs with acoustic levitation has emerged as a particularly promising strategy for non-contact manipulation and reaction initiation [18]. Zang et al. [19] demonstrated that a polytetrafluoroethylene (PTFE)-coated LM levitated within a standing acoustic field undergoes continuous shape transformation—from a quasi-spherical form to an oblate ellipsoid—as the sound intensity increases. This deformation is accompanied by the formation of cavities at the poles of the particle shell, resulting from particle migration toward the equatorial region driven by acoustic radiation pressure. Notably, this process is reversible: decreasing the sound intensity allows the cavity to heal, enabling the LM to serve as a switchable microreactor into which the reagents can be introduced and subsequently sealed. Expanding on this concept, Chen et al. [20] demonstrated that multiple LMs levitated acoustically are drawn together and coalesce into a single, larger marble, driven by pressure gradients on the marble surface generated by the acoustic field. This coalescence triggered an indicator reaction between the phenolphthalein and NaOH solutions contained in separate marbles, thereby demonstrating that acoustic levitation can controllably initiate chemical reactions between reagents isolated within individual LMs. Collectively, these studies establish acoustic levitation as a robust technique for the contactless manipulation of LMs and highlight its potential as a reaction platform for small-scale chemical and biological systems [18-20].

[a] Graduate School of Engineering, Kogakuin University, Tokyo, 163-8677, Japan.
[b] Marine Environment & Engine System Department, National Maritime Research Institute, Tokyo 181-0004, Japan
[c] Department of Mechanical Engineering, Kogakuin University, Tokyo, 163-8677, Japan. E-mail: kojihasegawa@cc.kogakuin.ac.jp

Acoustic levitation is a containerless processing technique that enables the levitation, transport, and mixing of samples through the acoustic radiation force generated by a standing ultrasonic wave [21, 22]. Compared with electrostatic or magnetic levitation techniques, acoustic levitation is applicable to a broad spectrum of materials, regardless of their electromagnetic properties [23]. However, because levitation is achieved by balancing the acoustic radiation pressure with the weight of the sample, levitated droplets often demonstrate nonlinear and complex behaviours, including vibration, significant interface deformation, internal and external flow [24-26], and evaporation [27, 28]. These phenomena can compromise the precision of non-contact fluid control and impact sample integrity, highlighting the importance of a detailed understanding of droplet dynamics [29, 30]. Evaporation, in particular, is crucial in applications such as drug discovery and analytical chemistry. However, the evaporation behaviour of LMs levitated in an acoustic field remains insufficiently characterised. Specifically, the mechanisms by which the hydrophobic particle shell influences evaporation and condensation[31], as well as the effects of acoustic streaming on mass transfer around the droplet [32], have yet to be systematically explored [22].

Accordingly, this study aims to identify the evaporation dynamics of LMs levitated in an acoustic field by visualising and quantitatively evaluating droplet shape evolution and the surrounding density gradients using backlight and Background-oriented Schlieren (BOS) method [33-40].

## 2. Experimental design and evaporation model

### 2.1 Experimental setup

A schematic of the experimental setup utilised in this study is shown in Fig. 1 (a). A sine wave signal was generated using a function generator (Agilent Technologies Japan 33511 B), subsequently amplified using an amplifier (NF Corporation 4502), and then input to an ultrasonic transducer (NGK Spark Plug Co. D4520PC). The input voltage and amplified power were monitored in real time using a power meter (Yokogawa Test & Measurement Corporation, WT310-DC1). Ultrasonic waves emitted from the bottom horn, which was coupled to the ultrasonic transducer, were reflected by a top reflector positioned at a distance corresponding to an integer multiple of half the wavelength. This configuration established an acoustic standing wave between the horn and reflector, as shown in Fig. 1 (b) [41]. To achieve droplet levitation, a liquid droplet was manually introduced near the acoustic pressure node of the acoustic standing wave using a syringe and needle. Furthermore, the acoustic levitator was enclosed in the chamber box shown in Fig. 1 (a) to control humidity. A pipe mounted on the side of the box was connected to a drying unit consisting of an air pump and a silica gel desiccant. The pump (ULVAC DA-40S) drew in air, passed it through a container packed with silica gel, and the desiccant adsorbed moisture from the air. Consequently, dry air could be continuously supplied to the chamber box. Visualisation of the levitated droplets was accomplished with a high-speed camera (Photron, FASTCAM Mini AX200) equipped with backlight illumination. The captured images were processed and analysed using a computer and in-house code (MATLAB R2023a). The reflector utilised in the experiment featured a radius of curvature of 36 mm (*R*36). The operating frequency was 20.3 kHz, corresponding to a wavelength λ of 18 mm. Both the horn and reflector had a width of 36 mm (=2λ). The separation distance between the horn and reflector was 48 mm (≈5λ/2), and droplets were levitated near the third pressure node (≈3λ/2 from the bottom reflector). All experiments were conducted at room temperature (10 to 30 °C) and humidity (10 to 90%). The test samples consisted of water droplets with initial equivalent diameters ($d_0$) ranging from 2.2 to 3.2 mm. The initial aspect ratio varied between 1.0 and 2.8. The aspect ratio, defined as $As = b/a$, represents the ratio of the equatorial (major) diameter ($b$) to the polar (minor) diameter ($a$) as represented in Fig. 1 (c). Here, using $b$ and $a$, the equivalent diameter of a levitating droplet with an oblate spheroidal shape is also defined as $d = \sqrt[3]{ab^2}$. The droplet diameters and aspect ratios were quantified using MATLAB image analysis. High-speed imaging was conducted at 50 fps with a resolution of 512 × 512 pixels, whereas those of the infrared camera (FLIR Systems, A6750sc, MWIR) operated at 1 fps with the same resolution. The emissivity of the infrared (IR) camera was determined by comparing it with a contact thermometer. The temperature of water in a beaker was measured while the emissivity was varied, and good agreement was obtained between the IR camera and the contact thermometer over the emissivity range of 0.96 to 1.00. Based on these results, an emissivity of 0.96 was adopted to measure the temperature of the LM surface, assuming the liquid inside the LM was water. The ambient temperature and RH were monitored using a high-precision thermometer and a hygrometer (Testo, testo622).

The uncertainty in droplet diameter was <1.4%, because at $d_0$ = 2.2 mm (the smallest diameter in the present study), the standard deviation with three measurements was <2 pixels, with a spatial resolution of ~15 μm/pixel. Meanwhile, the uncertainty of the droplet surface temperature was ±2 °C with the IR camera used in the present study.

### 2.2 Preparation of LMs

The LMs were prepared by gently depositing a droplet onto a bed of hydrophobic particles using a syringe, allowing the particles to adhere to the interface of the droplet and encapsulate it. In this study, the hydrophobic particles used were Lycopodium (LYCO) spores derived from LYCO clavatum (Sigma-Aldrich, St. Louis, MO, USA). As shown in Fig. 2, each droplet was placed on the particle layer and gently rolled back and forth approximately five times to ensure uniform coating. The average particle diameters were approximately 30 μm [42]. For comparison, additional experiments were conducted using LMs prepared with PTFE particles (Sigma-Aldrich, St. Louis, MO, USA) and non-sticking water droplets stabilised by a particulate polydimethylsiloxane organogel (PG) [43]. The typical snapshots of each LM with the microscopic images of the LMs surfaces and physical properties are shown in Fig. 3 and Table 1.

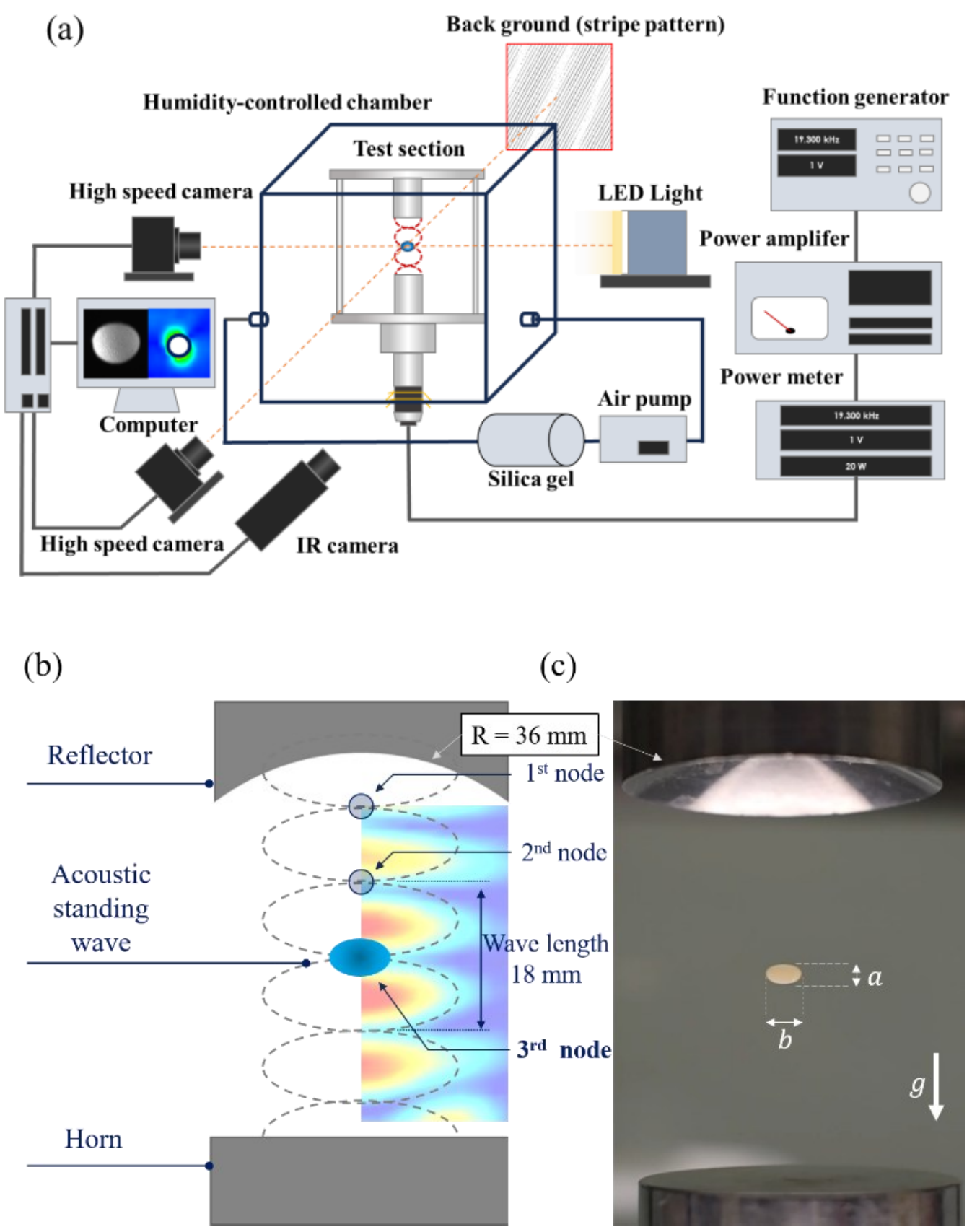

Fig. 1 Schematic of the acoustic levitator: (a) experimental setup, (b) levitation principle, and (c) levitation of the LM.

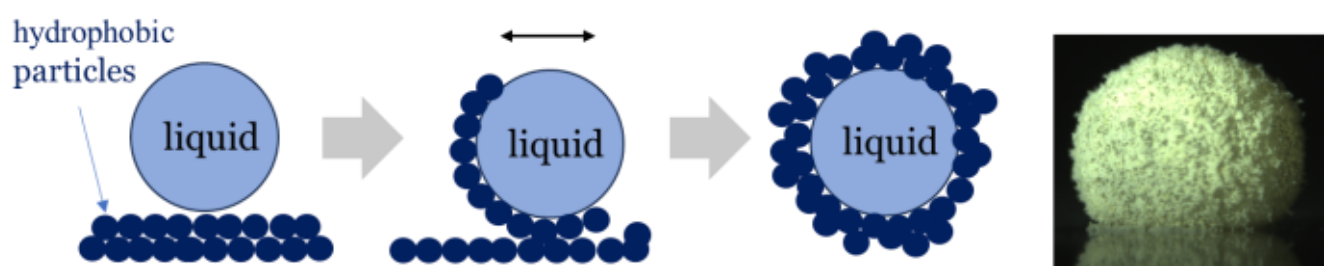

Fig. 2 Preparation of LM. Snapshot represents the Lycopodium–based LM.

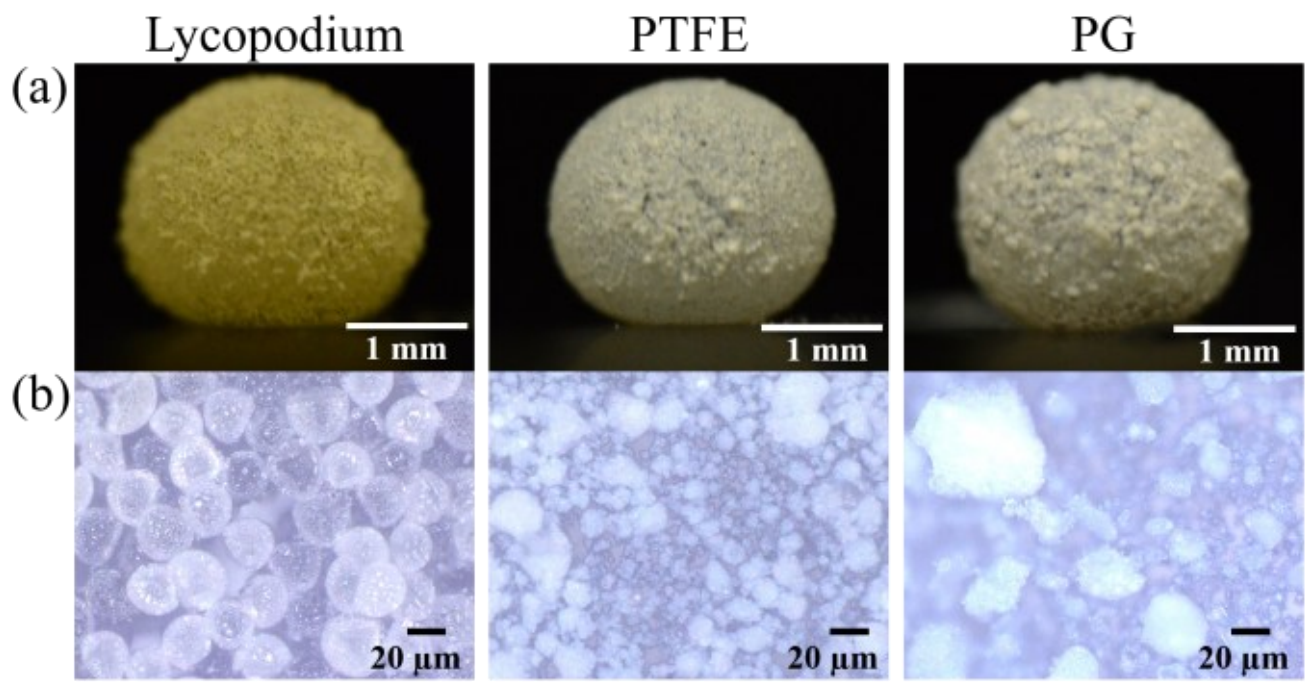

Fig. 3 Prepared different liquid marbles: (a) Snapshots of LMs and (b) microscopic images of LM surfaces.

Table 1 Physical properties of hydrophobic particles [42, 43]

| Sample | Average particle size [μm] | Density [g/cm$^3$] |
|---|---|---|
| LYCO | 30 | 1.05 |
| PTFE | 1 | 2.15 |
| PG | 90 | N/A |

### 2.3 Image analysis: Background-oriented Schlieren

To visualise the density gradient distribution around the droplets, the BOS method was employed. BOS is an optical measurement technique that leverages variations in the refractive index of a fluid, which occur in response to disturbances within the flow field, thereby enabling the visualisation of density fluctuations. As shown in Fig. 4, the BOS technique involves positioning a background image with an arbitrary pattern behind the target flow field and capturing the scene using a digital imaging system. Distortions in the background pattern, caused by density gradients in the flow field, are then analysed to reveal the flow structure and associated density variations.

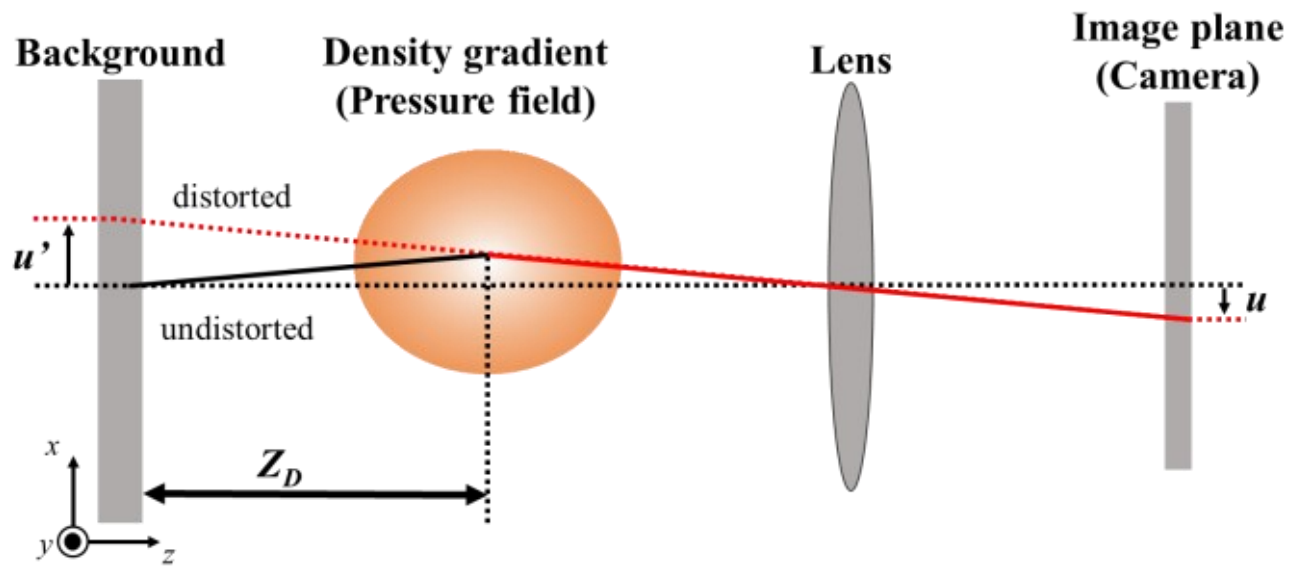

Fig. 4 Principle of the BOS Method.

In the BOS method, as shown in Fig. 4, the background and target objects were aligned along a common optical axis. Background images were acquired both with and without the target object present, and the resulting displacement field is determined using a displacement detection algorithm (Seika Digital Image Co., Ltd.). In the absence of a target, the camera records the light path as indicated by the dashed black line. By contrast, when the target was present, changes in the refractive index of the surrounding fluid caused the light to bend, and the camera detected this along the path indicated by the solid red line. The apparent distortion observed in the background image or image sensor, denoted as $u$ or $u'$, is interpreted as the displacement field. Specifically, the displacement in the x-direction, $u$ observed in the captured image is proportional to the line-of-sight integral of the density gradient distribution and can be expressed as follows [44-46]:

$$u = \frac{Z_D}{n_0}\int_{Z_u-\Delta Z_D}^{Z_D+\Delta Z_D}\frac{\partial n}{\partial x}dz = \frac{KZ_D}{n_0}\int\frac{\partial \rho}{\partial x}dz \tag{1}$$

where $n$ represents the refractive index, $n_0$ represents the refractive index of water, $Z_D$ represents the distance from the background to the target object, $\Delta Z_D$ represents the half-width of the target, and $K$ represents the Gladstone–Dale constant. The flow-density field was calculated from the measured displacement.

**2.4 Evaporation model**

Accurate estimation of droplet evaporation is particularly critical for non-contact manipulation. Therefore, theoretical predictions were compared with experimental results. The $d^2$-law, which describes the mass transport of single-component droplets, is widely employed to estimate the evaporation of such droplets and is expressed as follows: [47, 48]

$$\left(\frac{d}{d_0}\right)^2 = 1 - \frac{8DM}{\rho_l R}\left(\frac{P_{\mathrm{sur}}}{T_{\mathrm{sur}}} - \frac{P_\infty}{T_\infty}\times\frac{RH}{100}\right)\frac{t}{d_0^2} \tag{2}$$

where $D$ represents the diffusion coefficient, $M$ represents the molecular weight, $\rho_l$ represents the density, $R$ represents the gas constant, $P$ represents the vapour pressure, $T$ represents the temperature, $RH$ represents the relative humidity, $t$ represents the time, and $d_0$ represents the initial equivalent diameter of the droplet. The subscript sur represents the droplet surface and $\infty$ represents the ambient gas.

## 3. Results and discussion

**3.1 Evaporation process of water droplet and LM**

The evaporation process of a water droplet and water-containing LM (LYCO) is shown in Fig. 6. The corresponding time evolution of the aspect ratio ($As$) for each droplet is presented in Fig. 6 (a). As illustrated, the LM flattened over time, and a droplet not encapsulated as an LM is subject to the vertical sound pressure when levitated in an acoustic field. Larger droplets tend to assume an ellipsoidal shape. As evaporation proceeds and the droplet size decreases, the relative influence of surface tension increases, causing the droplet to become more spherical. Conversely, Fig. 6 (a) shows that $As$ of the LM increased from 1.4 to 1.9 between 0 and 300 s. The coated particles of the LM suppress the reduction in surface area by maintaining contact and mutual interference among the particles as the internal droplet volume decreases owing to evaporation. Consequently, the surface area remains largely unchanged, and the LM flattens to accommodate the reduction in the internal droplet volume. Furthermore, the relative increase in the influence of vertical sound pressure as the internal volume decreases is considered a contributing factor to the observed flattening.

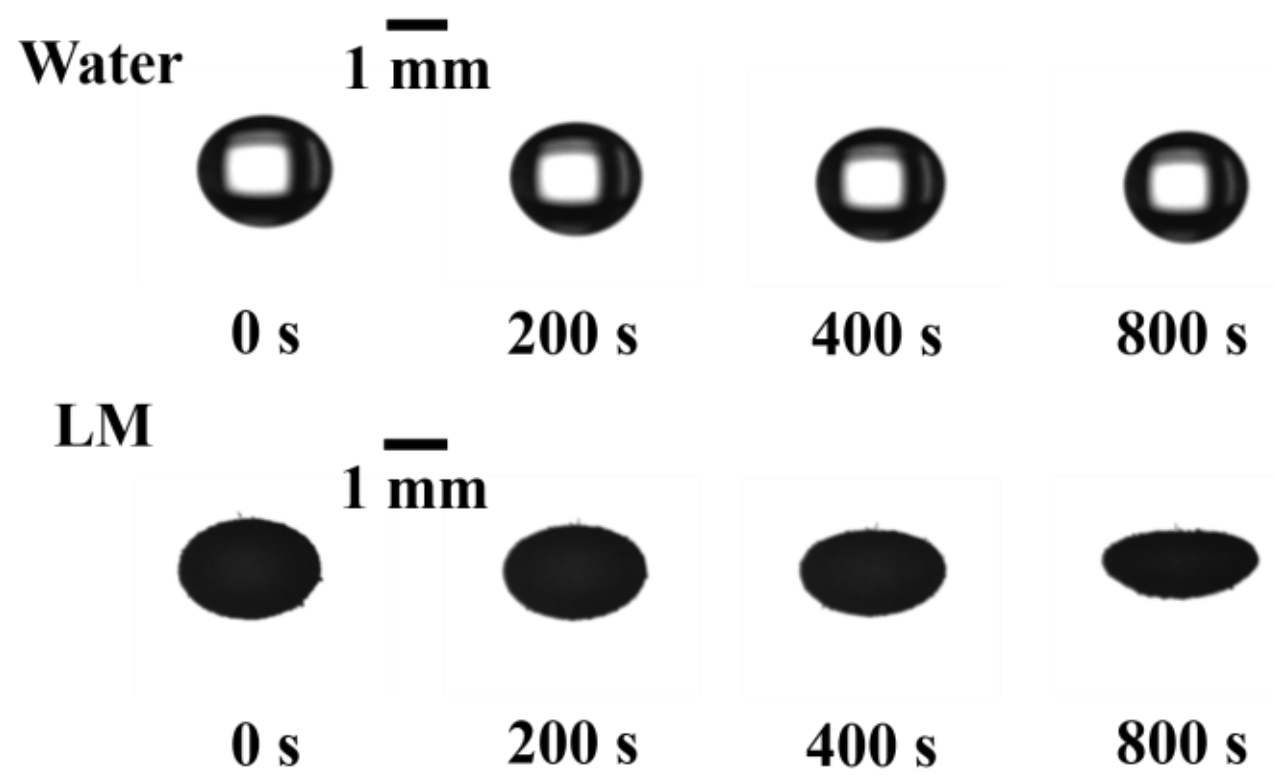

Fig. 5 Snapshots of (top) the water droplet and (bottom) LM (LYCO) during evaporation in acoustic levitation.

The time evolution of the normalised surface area $(d/d_0)^2$ of the levitating water droplet and an LM (LYCO) containing water is shown in Fig. 6 (b). The vertical axis represents the square of the droplet diameter $d$ normalised by the square of the initial droplet diameter $d_0$. As illustrated, no discernible difference in the evaporation rates between pure water droplets and LM-encapsulated water droplets was observed. This similarity is attributed to the ability of the LM to suppress both internal evaporation and condensation, resulting in a net evaporation rate comparable with that of the external droplet.

Given the absence of a significant difference in evaporation rates, the evaporation behaviour of the LMs was further analysed using the theoretical model described by Eq. (2). The droplet surface temperature was determined through interfacial temperature measurements obtained via infrared thermography, whereas the diffusion coefficient and density of the levitated sample were estimated based on the known physical properties of water. A comparison between the experimental data and predictions based on the $d^2$-law is shown in Fig. 6 (b). The red line represents the predicted values derived from the $d^2$-law. The results demonstrate that the experimental evaporation rates of LM water droplets closely align with the theoretical predictions over the first 200 s. These findings suggest that the evaporation process of LM water droplets can be accurately described using the $d^2$-law, originally formulated for conventional water droplets.

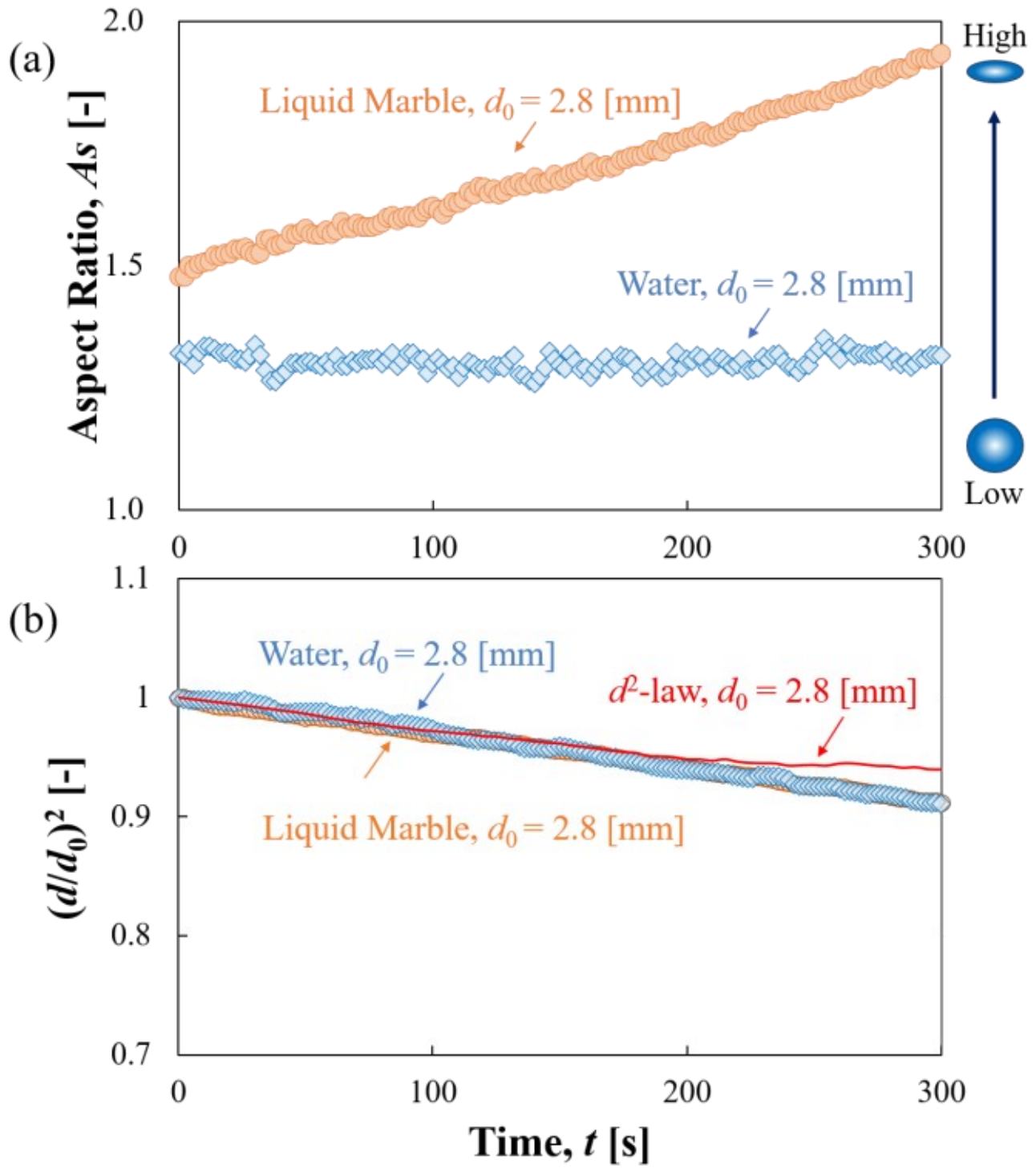

Fig. 6 Evaporation process of the water droplet and LM (LYCO): (a) aspect ratio and (b) the normalised surface area.

### 3.2 Density gradient distribution around water droplet and LM

This section presents the results of the density gradient distribution measurements around the droplets obtained using the BOS method. The density gradient surrounding the water droplets (a) and LM (b) is shown in Fig. 7. The colour bars represent deviations from the reference pattern image. Furthermore, schematics of the levitated droplets captured simultaneously were overlaid on the Schlieren images to enhance visibility. These results enabled visualisation of the density gradients—indicative of concentration changes in the surrounding gas phase owing to evaporation—generated around the LM levitated in the acoustic field. Notably, a significant density gradient was observed around the LM-coated droplets. No significant suppression of droplet evaporation by hydrophobic particles distributed at the interface was observed. These results were consistent with those presented in the previous section.

Significant density gradients were primarily observed at the upper left and lower right regions of the droplet interface for all the droplets. Evaporation was most significant at the top and bottom of the droplet, and less so at the sides. This distribution is attributed to the flow field generated around the droplet, which is characteristic of the acoustic field [49]. The flow field surrounding a droplet levitated in an acoustic field is believed to consist of an acoustic flow near the boundary and vortices formed outside this region. Assuming that the LM-coated water droplet formed a similar flow field, the vapour generated at the sides of the droplet was collected at the top or bottom by the surrounding flow field.

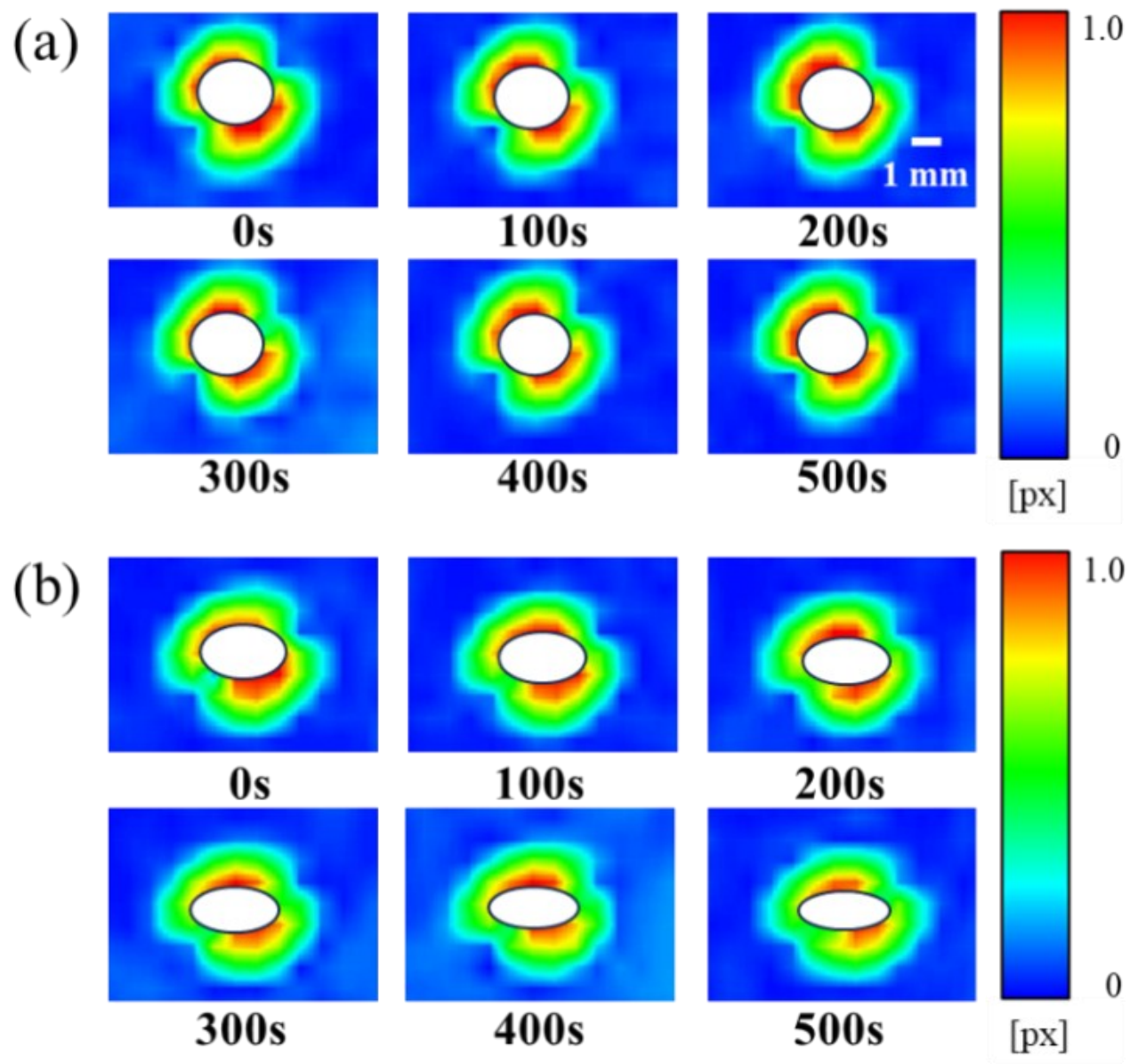

Fig. 7 BOS measurement results for density gradient surrounding (a) the water droplet and (b) LM (LYCO).

### 3.3 Effect of initial droplet size

We evaluated the evaporation process by varying the diameter of LM-coated water droplets. The experiment was conducted at a controlled room temperature of 25 ± 2 °C and an ambient humidity of 50 ± 10%. The evaporation process for different initial droplet diameters is shown in Fig. 8. The initial diameters ($d_0$) of the LM droplet were set to 2.2 mm, 2.8 mm, and 3.2 mm. Over 400 s, the value of $(d/d_0)^2$ decreased by 10% for $d_0$ = 3.2 mm and by 20% for $d_0$ = 2.2 mm, indicating a higher evaporation rate for smaller initial droplet diameters. Assuming the droplets are spherical, the surface area-to-volume ratio is expressed as $S/V = (4\pi r^2)/(4\pi r^3/3) = 6/d$, demonstrating that smaller droplets have a greater surface area per unit volume. Consequently, the evaporation rate increased as the droplet diameter decreased.

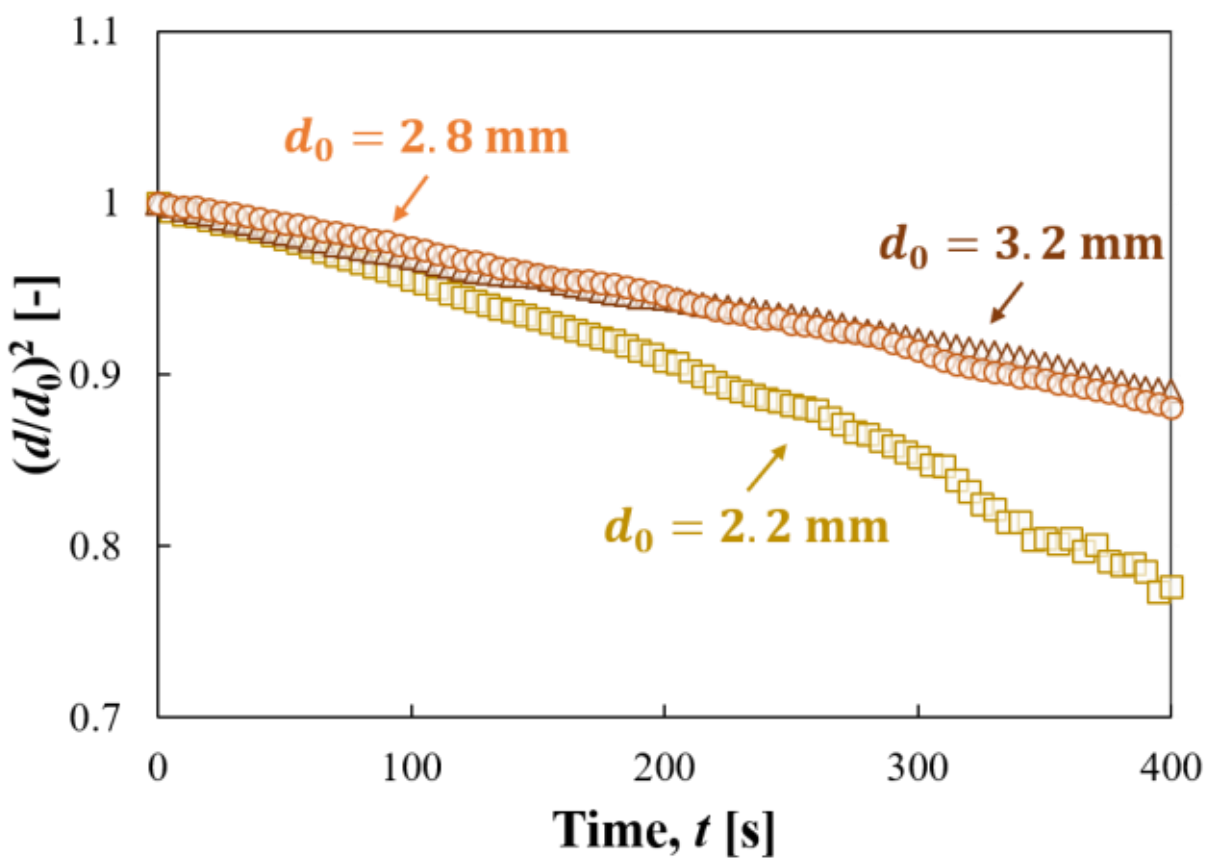

Fig. 8 Effect of the initial LMs diameter on evaporation process.

### 3.4 Effect of relative humidity

The correlation between the evaporation rates of the LM and water droplets, as well as the relative humidity, is shown in Fig. 9. The vertical axis represents the slope (β) of the evaporation curve defined using Eq. (3), where the horizontal axis represents the RH.

$$\left(\frac{d}{d_0}\right)^2 = 1 - \beta t \quad (3)$$

To explore the effect of particle size, the evaporation results for LM prepared with hydrophobic particles of different sizes are also presented. The schematic of the evaporation behaviour of a water droplet and an LM under high- and low-humidity conditions based on the results in Fig. 9 is shown in Fig. 10.

As shown in Fig. 9, at an RH of approximately 40%, where evaporation behaviour adheres to the $d^2$-law, only minor differences were observed among the plots, and all samples demonstrated comparable evaporation rates. By contrast, under low-humidity conditions (approximately 20% RH), water droplets evaporated more rapidly than LMs. This phenomenon is attributed to the presence of hydrophobic particles at the interface, which impede the evaporation of internal droplets and restrict mass transfer, as shown in Fig. 10 (a). In the LMs, both evaporation (red arrows) and condensation (green lines) were suppressed by the particle layer. However, under low-humidity conditions, evaporation was significantly accelerated for both LMs and water droplets, resulting in a significant difference in their evaporation behaviour.

Conversely, under high-humidity conditions (RH > 40%), the trend reversed: LMs displayed a faster evaporation rate than the water droplets. As shown in Fig. 10 (b), this can be attributed to the suppression of condensation in the LMs by the hydrophobic particle layer at the interface. In high-humidity environments, condensation is significantly enhanced by both LMs and water droplets, making the differences in the condensation process more evident.

Therefore, in low-humidity environments, where evaporation predominates, the evaporation-suppressing effect of the LM particle layer becomes dominant. This resulted in a slower evaporation rate for the LMs than for the water droplets. Conversely, under high-humidity conditions, where condensation predominates, the suppression of condensation by the particle layer leads to a higher apparent evaporation rate for the LMs than for the water droplets.

Comparative analysis of the evaporation rates among different particle types revealed that PG, characterised by its notably large particle size, evaporates rapidly under both low- and high-humidity conditions. This accelerated evaporation is attributed to the unique properties of LMs, where larger particles tend to form monolayer structures. Such structures impose fewer restrictions on mass transfer compared with PTFE and LYCO, which typically form multilayer structures [50].

These findings suggest that the evaporation and condensation behaviour of LMs can be modulated by varying both the RH and particle size.

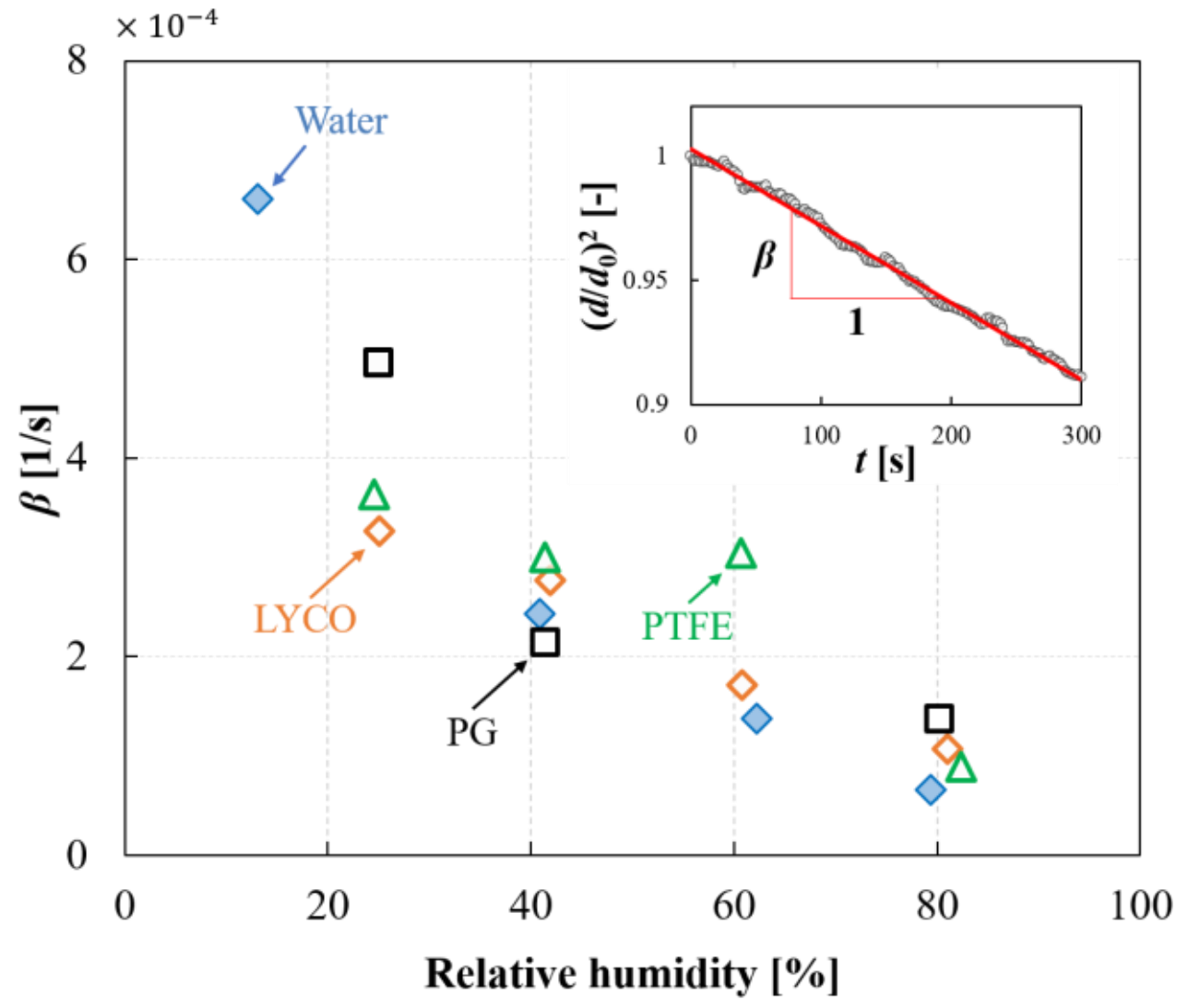

Fig. 9 Relationship between the evaporation rate ($\beta$) and RH with LMs (LYCO, PTFE, and PG) and water droplets for $d$ = 2.5± 0.1 mm.

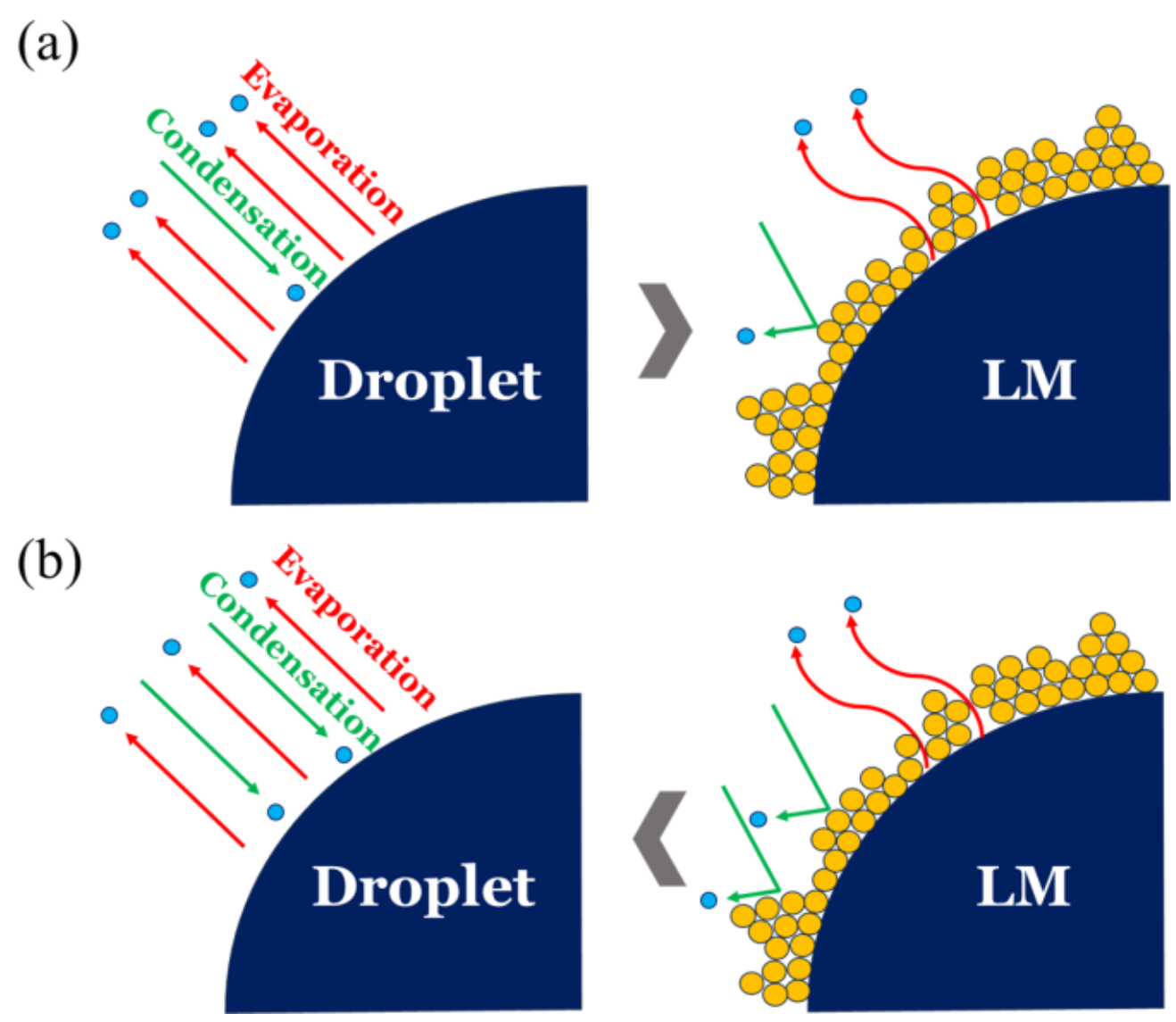

Fig. 10 Schematics of the phase-change process of water droplets and LMs for (a) lower and (b) higher relative humidity. For clarity, the figure is not presented to scale.

## 4. Conclusion

In this study, we examined the evaporation dynamics of LMs levitated in an acoustic field, employing a combination of visualisation techniques and quantitative measurements. Our findings demonstrated that the presence of hydrophobic particles at the LM interface influenced droplet morphology during evaporation, resulting in a characteristic flattening effect. This can be attributed to the suppression of surface area reduction and the relative enhancement of acoustic radiation effects. Despite these shape changes, the net evaporation rate of LM water droplets did not differ significantly from that of pure water droplets, indicating that the evaporation process of the LM droplets can be accurately described using the classical $d^2$-law.

Density gradient measurements using the BOS method revealed distinct flow patterns and vapour distributions around the LM and water droplets, highlighting the significant influence of the surrounding acoustic field on local evaporation behaviour. Additionally, quantitative analysis of the initial droplet diameter confirmed that smaller droplets evaporated more rapidly, owing to their higher surface area-to-volume ratios. Furthermore, the evaporation behaviour strongly depended on both RH and particle size. Under low-humidity conditions, the particle layer suppressed evaporation; however, under high-humidity conditions, it inhibited condensation, resulting in a faster evaporation rate than that for water droplets. Large particles forming monolayer structures facilitated mass transfer, thereby enhancing the evaporation rate relative to multilayered, particle-coated LMs.

Collectively, these findings provide a comprehensive understanding of the influence of droplet geometry, particle coating, acoustic levitation, and ambient conditions on LM evaporation. The ability to predict and control LM evaporation through the particle size and RH is valuable for developing lab-in-a-drop applications, such as micro-reactors, analytical chemistry, and droplet-based chemical or biological assays. This study provides essential insights for the design of practical, containerless microfluidic systems with tunable evaporation and condensation kinetics.

## Author contributions

K. N.: investigation, methodology, formal analysis, visualisation, writing – original draft. X. M.: visualisation, validation, writing – review & editing. K. H.: conceptualisation, supervision, resources, project administration, writing – review & editing.

## Conflicts of interest

There are no conflicts to declare.

## Data availability

All data supporting the findings of this study are included within the article, figures, and tables.

## Acknowledgements

This work was supported by JSPS KAKENHI Grant Numbers 23K17732 and 25K01162. The authors gratefully acknowledge Dr. Mizuki Tenjimbayashi for insightful discussions and for providing the PG.

## References

1. P. Aussillous and D. Quéré, *Nature*, 2001, 411, 924–927.
2. G. McHale and M. I. Newton, *Soft Matter*, 2011, 7, 5473–5481.
3. E. Bormashenko, *Soft Matter*, 2012, 8, 11018–11021.
4. E. Bormashenko, *Langmuir*, 2017, 33, 663–669.
5. K. R. Sreejith, C. H. Ooi, J. Jin, D. V. Dao and N.-T. Nguyen, *Lab Chip*, 2018, 18, 3770–3779.
6. K. P. Hapgood and B. Khanmohammadi, *Powder Technol*., 2009, 189, 253–262.
7. E. Bormashenko, R. Pogreb and A. Musin, *J. Colloid Interface Sci.*, 2012, 366, 196–199.
8. E. Bormashenko, A. Musin, G. Whyman and M. Zinigrad, *Langmuir*, 2012, 28, 3460–3464.
9. N. Nguyen, C. H. Ooi, P. Singha, J. Jin, K. R. Sreejith, H.-P. Phan and N.-T. Nguyen, *Processes*, 2020, 8, 793.
10. T. Arbatan, L. Li, J. Tian and W. Shen, Adv. Healthcare Mater., 2011, 1, 80–83.
11. T. Arbatan, A. Al-Abboodi, F. Sarvi, P. P. Y. Chan and W. Shen, *Adv. Healthcare Mater*., 2012, 1, 467–469.
12. F. Sarvi, K. Jain, T. Arbatan, P. J. Verma, K. Hourigan, M. C. Thompson, W. Shen and P. P. Y. Chan, *Adv. Healthcare Mater*., 2015, 4, 77–86.
13. R. K. Vadivelu, H. Kamble, A. Munaz and N.-T. Nguyen, *Sci. Rep*., 2017, 7, 12388.
14. F. Sarvi, T. Arbatan, P. P. Y. Chan and W. Shen, *RSC Adv*., 2013, 3, 14501–14508.
15. R. K. Vadivelu, C. H. Ooi, R.-Q. Yao, J. T. Velasquez, E. Pastrana, J. Diaz-Nido, F. Lim, A. K. Ekberg, N. -T. Nguyen and J. A. St John, *Sci. Rep*., 2015, 5, 15083.
16. J. Tian, T. Arbatan, X. Li and W. Shen, *Chem. Commun.,* 2010, 46, 4734–4736.
17. J. Tian, T. Arbatan, X. Li and W. Shen, *Chem. Eng. J.*, 2010, 165, 347–353.
18. A. Vashi, A. S. Yadav, N. Kashaninejad, and N. T. Nguyen, *Colloids Surf. A: Physicochem. Eng. Asp.*, 2025, 711, 136410.
19. D. Zang, J. Li, Z. Chen, Z. Zhai, X. Geng and B. P. Binks, *Langmuir*, 2015, 31, 11502-11507.
20. Z. Chen, D. Zang, L. Zhao, M. Qu, X. Li, X. Li, L. Li and X. Geng, *Langmuir,* 2017, 33, 6232–6239.
21. A. Marzo, A. Seah, B. W. Drinkwater, D. R. Sahoo, B. Long and S. Subramanian, *Nat. Commun*., 2015, 6, 8661.
22. H. Chen, Z. Hong, and D. Zang, *Droplet*, 2024, 3, e95.
23. E. H. Brandt, *Science*, 1989, 243, 349–355.
24. K. Hasegawa, Y. Abe, A. Kaneko and K. Aoki, Int. J. Transp. Phenom., 2011, 12, 151–160.
25. K. Hasegawa, Y. Abe and A. Goda, *npj Microgravity*, 2016, 2, 16004.
26. A. Watanabe, K. Hasegawa and Y. Abe, *Sci. Rep*., 2018, 8, 10221.
27. Y. Niimura and K. Hasegawa, *PLoS One*, 2019, 14, e0212074.
28. M. Mitsuno and K. Hasegawa, *Phys. Fluids*, 2024, 36, 033328.
29. D. Foresti, M. Nabavi, M. Klingauf, A. Ferrari and D. Poulikakos, *Proc. Natl. Acad. Sci. U. S. A*., 2013, 110, 12549–12554.
30. R. H. Morris, E. R. Dye, P. Docker, and M. I. Newton, *Phys. Fluids*, 2019, 31, 101301.
31. K. R. Sreejith, C. H. Ooi, D. V. Dao and N.-T. Nguyen, RSC Adv., 2018, 8, 15436–15443.
32. Y. Sasaki, K. Hasegawa, A. Kaneko, and Y. Abe, *Phys. Fluids*, 2020, 32, 072102.
33. G. S. Settles, *Schlieren and Shadowgraph Techniques: Visualizing Phenomena in Transparent Media,* Springer, Berlin, 2001.
34. H. Richard and M. Raffel, *Meas. Sci. Technol.*, 2001, 12, 1576–1585.

35. G. E. A. Meier, *Exp. Fluids*, 2002, 33, 181–187.
36. G. E. Elsinga, B. W. van Oudheusden, F. Scarano and D. W. Watt, *Exp. Fluids*, 2004, 36, 309–325.
37. L. Venkatakrishnan and P. Suriyanarayanan, *Exp. Fluids*, 2009, 47, 463–473.
38. M. Raffel, *Exp. Fluids*, 2015, 56, 60.
39. F. Nicolas, V. Todoroff, A. Plyer, G. Le Besnerais, D. Donjat, F. Micheli, F. Champagnat, P. Cornic and Y. Le Sant, *Exp. Fluids*, 2016, 57, 13.
40. P. Zhang, Z. Xu, T. Wang and Z. Che, *Int. J. Heat Mass Transfer*, 2021, 168, 120880.
41. M. Mitsuno, X. Ma and K. Hasegawa, *Phys. Fluids*, 2025, 37, 084107.
42. B. Renganathan, A. Viswanathan, D. Sastikumar and S. S. Hameed, *Frontiers in Optics 2007*, Paper No. LThB5, 2007.
43. M. Tenjimbayashi and R. Tamate, *J. Mater. Chem. A*, 2024, 12, 16343–16349.
44. L. Venkatakrishnan and G. E. A. Meier, *Exp. Fluids*, 2004, 37, 237–247.
45. S. Ichihara, T. Shimazaki and Y. Tagawa, *Exp. Fluids*, 2022, 63, 182.
46. K. Hayasaka and Y. Tagawa, *Exp. Fluids*, 2019, 60, 171.
47. A. L. Yarin, G. Brenn, O. Kastner, D. Rensink and C. Tropea, *J. Fluid Mech*., 1999, 399, 151–204.
48. Y. Maruyama and K. Hasegawa, *RSC Adv*., 2020, 10, 1870–1877.
49. Y. Sasaki, K. Kobayashi, K. Hasegawa, A. Kaneko and Y. Abe, *Phys. Fluids*, 2019, 31, 102109.
50. T. H. Nguyen, K. Hapgood and W. Shen, *Chem. Eng. J.*, 2010, 162, 396–405.